\documentclass[pra,twocolumn,superscriptaddress,showpacs,amsmath,amssymb,amsfonts]{revtex4-2}
\usepackage{graphicx}
\usepackage{xcolor}
\usepackage{subfigure}
\usepackage{bm}
\usepackage{siunitx}
\usepackage{amsthm,mathtools}
\usepackage[normalem]{ulem}

\def\U#1{{\rm #1}} 
\def\u#1{_{\rm #1}}

\newcommand{\vect}[1]{\boldsymbol{#1}}
\newcommand{\od}[2]{\frac{\mathrm{d} #1}{\mathrm{d} #2}}

\begin{document}
\title{
  All-optical control of coherent perfect absorption via frequency conversion
}
\author{Rikizo Ikuta}
\affiliation{
  Graduate School of Engineering Science, The University of Osaka, 
  Toyonaka, Osaka 560-8531, Japan}
\affiliation{
  Center for Quantum Information and Quantum Biology, The University of Osaka, 
  Toyonaka, Osaka 560-0043, Japan}
\author{Hirokazu Kobayashi}
\affiliation{
  Graduate School of Engineering, Kochi University of Technology, Kochi 782-8502, Japan
  }
\author{Hiroki Takahashi}
\affiliation{
  Experimental Quantum Information Physics Unit,
  Okinawa Institute of Science and Technology Graduate University,
  Okinawa 904-0495, Japan
}

\begin{abstract}
  Coherent perfect absorption~(CPA) extinguishes optical fields through interference and dissipation.
  In conventional implementations,
  the required dissipation is typically provided by material loss that is largely fixed after fabrication.
  Here we demonstrate all-optically controllable CPA based on frequency conversion 
  in a periodically poled lithium niobate waveguide resonator. 
  Under a strong pump, the nonlinear frequency-conversion process acts
  as an effective beamsplitter-like linear coupling between a resonant signal field at \SI{1581}{nm}
  and a non-resonant frequency-converted mode at \SI{780}{nm},
  creating a dynamically tunable effective loss channel, without relying on intrinsic material absorption.
  With coherent two-sided signal injection, we observe up to \SI{92}{\%} absorption. 
  We further introduce environment-assisted CPA by injecting an external field into the frequency-converted environmental mode,
  turning the environment from a passive loss reservoir into an addressable coherent control port. 
\end{abstract}
\maketitle

\section{Introduction}
Coherent perfect absorption~(CPA)~\cite{Chong2010,Wan2011} is a fundamental wave phenomenon
that enables the complete extinction of coherent optical fields through the interplay of interference and dissipation. 
As the time-reversed counterpart of lasing, 
CPA has emerged as a versatile paradigm for manipulating light, 
with applications including all-optical switching~\cite{Zhang2012-2,Fang2014,Wang2023},
modulation~\cite{Rao2014,Fang2015,Xomalis2018}, and spectroscopy~\cite{Fang2016}.
It has also been extended to the quantum regime,
including single-photon absorption~\cite{Roger2015},
N00N-state absorption~\cite{Roger2016},
nonlocal control of single-photon absorption~\cite{Altuzarra2017},
and quantum interference at lossy beamsplitters~\cite{Vest2017}.

CPA has been demonstrated in various platforms,
including graphene films~\cite{Rao2014,Kakenov2016}, metasurfaces~\cite{Zhang2012-2,Urade2016}, and optical resonators~\cite{Wan2011,Soleymani2022}. 
In conventional CPA,
the required optical loss is typically provided by intrinsic material absorption or passive linear dissipation. 
Recent studies have explored actively tunable CPA 
by engineering either the effective loss~\cite{Kats2012,Kakenov2016,Zhang2019} or the scattering matrix~\cite{Krishna2026},
using external control parameters such as temperatures and electrical voltages. 
These approaches generally rely on modifying material, structural, or circuit parameters of the system. 
In contrast, all-optical control offers an attractive route toward ultrafast operation
and seamless integration with photonic networks.
More recently, nonlinear and optically driven CPA schemes based on parametric
interactions have also been demonstrated~\cite{Cui2024,Galiffi2026}. 
These schemes rely on phase-sensitive mixing of the signal with parametrically generated field.
In the quantum regime, such parametric interactions correspond to quadrature squeezing~\cite{Milburn1981,Lyubarov2022}, rather than to passive linear loss underlying the conventional CPA.

In this paper, we propose and experimentally demonstrate a platform for CPA 
based on optical frequency conversion within a two-sided Fabry-P\'{e}rot~(FP) cavity, 
which we term frequency-conversion-based CPA~(FC-CPA). 
Although FC-CPA is mediated by a nonlinear optical process,
frequency conversion driven by a classical pump realizes a photon-number-conserving,
beamsplitter-like linear interaction between the resonant signal mode of the two-sided cavity
and a spectrally distinct frequency-converted mode. 
For a supermode which is a symmetric superposition of two input modes~(described later in Sec.~\ref{sec:theory}),
this coupling acts as a pump-tunable loss,
retaining the linear-scattering character of conventional CPA while making its effective loss all-optically tunable.
This picture holds even in the quantum regime in contrast to the CPA using parametric interactions. 
Our system utilizes a periodically poled lithium niobate waveguide resonator~(PPLN/WR),
where an effective optical loss for the resonant signal mode at \SI{1581}{nm} is induced through its up-conversion
to a non-resonant frequency-converted environmental mode at \SI{780}{nm} via a strong pump field at \SI{1540}{nm}.

In addition, we introduce environment-assisted CPA,
in which an external field is simultaneously injected into the 780-nm environmental mode. 
By reinterpreting the environment channel not merely as a passive loss reservoir but as a coherent control port,
we show that the environmental input enables active enhancement or suppression of energy absorption,
even under non-ideal conditions such as off-resonance detuning or non-critical coupling. 
This versatile and actively controllable platform provides a new route
toward coherent control of dissipation in reconfigurable photonic interfaces. 

\section{Theoretical model}
\label{sec:theory}
\begin{figure}
 \begin{center}
      \scalebox{1}{\includegraphics{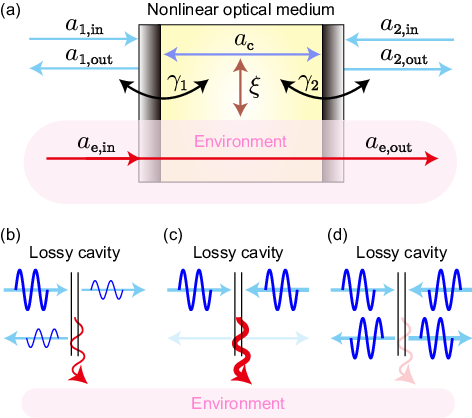}}
    \caption{
      Concept of FC-CPA, 
      based on a two-sided Fabry-P\'{e}rot cavity with frequency-conversion-induced optical loss.
      (a) Implementation of a lossy FP cavity using frequency conversion inside the cavity. 
      The signal light is injected into the lossy cavity acting as a BS
      from (b) one side, (c) both sides with identical phase,
      and (d) both sides with a $\pi$-phase difference between the inputs. 
    }
 \label{fig:concept}
 \end{center}
\end{figure}
We first explain the optical system for FC-CPA 
based on a two-sided FP cavity formed by a second-order nonlinear optical medium
with reflective optical coatings on its end facets. 
The conceptual design is shown in Fig.~\ref{fig:concept}~(a).
We note that in a strict sense, FC-CPA denotes the condition of perfect absorption. 
In the following, however, we also use the term to refer to its experimental implementation unless otherwise specified. 
We assume that the cavity mode is a single mode denoted by $a\u{c}$ 
at resonance angular frequency $\omega\u{c}$ 
and it is coupled to two external modes $a_1$ and $a_2$ at the left and right sides of the cavity with coupling constants 
$\sqrt{\gamma_1}$ and $\sqrt{\gamma_2}$, respectively.
This can be understood as a two-port BS~\cite{Ikuta2024} with inputs $a_{1,\U{in}}$ and $a_{2,\U{in}}$
and outputs $a_{1, \U{out}}$ and $a_{2,\U{out}}$ at a frequency $\omega$~(see Fig.~\ref{fig:concept}~(a)).

The optical loss of this BS is induced by frequency conversion 
between cavity mode $a\u{c}$ and a non-resonant mode $a\u{e}$ at frequency $\omega\u{e}$ as an environmental mode. 
We assume that the pump light at angular frequency $\omega\u{p}(=\omega\u{e} - \omega)$ 
for the frequency conversion is sufficiently strong. 
The time evolution for $a\u{c} = a\u{c}(t)$ 
in a frame rotating at the frequency $\omega$ is described by~\cite{Collett1984} 
\begin{align}
  \od{a\u{c}}{t} = \left( i\Delta\u{c} -\frac{\gamma + |\xi|^2}{2}\right) a\u{c}
  + \sum_{i=1,2}\sqrt{\gamma_i} a_{i, \U{in}} + \xi^* a\u{e,in},
  \label{eq:ac}
\end{align}
where $\Delta\u{c} = \omega - \omega\u{c}$ is the detuning
and $\gamma = \gamma_1 + \gamma_2 + \gamma\u{int}$ is the total loss of the cavity 
including an internal loss $\gamma\u{int}$~(not shown in the figure). 
$\xi$ is the effective coupling constant of the frequency conversion 
proportional to the complex amplitude of the pump light. 
From Eq.~\eqref{eq:ac}, we obtain the steady-state solution for $a\u{c}(\omega)$ as 
\begin{eqnarray}
  a\u{c}(\omega)
  = \frac{\sum_i \sqrt{\gamma_i} a_{i, \U{in}}(\omega) + \xi^* a\u{e,in}(\omega)}
      {\frac{1}{2}(\gamma+|\xi|^2)-i\Delta\u{c}}.  
\end{eqnarray} 
Using the input-output relations~\cite{Walls2007} as 
\begin{eqnarray}
  a_{1(2),\U{in}} + a_{1(2),\U{out}} &=& \sqrt{\gamma_{1(2)}} a\u{c},\\
  a\u{e,in} - a\u{e,out} &=& \xi a\u{c}, 
\end{eqnarray}
we derive the transformation matrix $T$ from $\vect{v}\u{in}=(a\u{1,in}, a\u{2,in}, a\u{e,in})^{\mathsf T}$ 
to $\vect{v}\u{out}=(a\u{1,out},a\u{2,out},a\u{e,out})^{\mathsf T}$ as $\vect{v}\u{out}=T\vect{v}\u{in}$. 
Assuming that $\gamma\u{int}=0$ and $\gamma_1=\gamma_2(=\gamma/2)$, for simplicity, 
$T$ is described by
\begin{eqnarray}
  T= A^{-1}
  \begin{pmatrix}
    -\frac{|C|^2}{2} + i\delta\u{c} & \frac{1}{2} & \frac{C^*}{\sqrt{2}}\\
    \frac{1}{2} & -\frac{|C|^2}{2} + i\delta\u{c} & \frac{C^*}{\sqrt{2}}\\
    -\frac{C}{\sqrt{2}} & -\frac{C}{\sqrt{2}} & 
    \frac{1 - |C|^2}{2} - i \delta\u{c}
  \end{pmatrix},
  \label{eq:T}
\end{eqnarray}
where $C=\xi/\sqrt{\gamma}$, $\delta\u{c}=\Delta\u{c}/\gamma$, and $A=\frac{1}{2}(1+|C|^2)-i\delta\u{c}$.
The above linear coupled-mode and input-output equations in Eqs.~(\ref{eq:ac}) -- (\ref{eq:T}) 
can be interpreted either as equations for annihilation operators or as equations for classical field amplitudes. 
In the following, we treat $a_{1,\U{in/out}}$, $a_{2,\U{in/out}}$ and $a\u{e,\U{in/out}}$
as classical complex amplitudes of the input and output traveling fields.

When there is no input from the environment as $a\u{e,in}=0$,
the scattering matrix $S$ from $\vect{u}\u{in}=(a\u{1,in},a\u{2,in})^{\mathsf T}$ to $\vect{u}\u{out}=(a\u{1,out},a\u{2,out})^{\mathsf T}$
satisfying $\vect{u}\u{out}=S\vect{u}\u{in}$ is described by a submatrix of $T$ as 
\begin{eqnarray}
  S= A^{-1}
  \begin{pmatrix}
    -\frac{|C|^2}{2} + i\delta\u{c} & \frac{1}{2} \\
    \frac{1}{2} & -\frac{|C|^2}{2} + i\delta\u{c} 
  \end{pmatrix}.
  \label{eq:S}
\end{eqnarray}
The eigenvalues and eigenvectors of $S$ are given by
\begin{eqnarray}
    \begin{cases}
      \lambda_+ = -1 + A^{-1} & \U{for}\ \vect{u}_{+}=(1,1)^{\mathsf T}/\sqrt{2}\\
      \lambda_- = -1 & \U{for}\ \vect{u}_{-}=(1,-1)^{\mathsf T}/\sqrt{2}. 
  \end{cases}
\label{eq:ev}
\end{eqnarray}
These two eigenvectors correspond to the symmetric and antisymmetric supermodes of the two input channels, respectively. 
The symmetric and antisymmetric inputs are written as $\vect{u}\u{in}=\alpha\u{s,in}\vect{u}_\pm$ 
using an overall complex amplitude $\alpha\u{s,in}$.
Since the symmetric input couples to the frequency-conversion-induced loss channel,
its eigenvalue depends on $C$ and $\delta\u{c}$. 
By contrast, the antisymmetric input remains decoupled from the loss channel,
and therefore no dissipation into the environment is induced regardless of $C$ and $\delta\u{c}$. 

From the perspective of FC-CPA, we are interested in the symmetric supermode. 
When the pump light for frequency conversion is turned off, $C=0$ is satisfied. 
In this case, we obtain $|\lambda_+|=1$, since no optical loss is induced. 
On the other hand, when frequency conversion is induced by the pump light, the process reduces $|\lambda_+|$, 
and eventually yields $\lambda_+ = 0$ for $|C|=1$ under the resonant condition $\delta\u{c}=0$. 
At this critical point, for a single-sided input from either the left or right side,
the cavity acts as a lossy BS, in which half of the incident photons are converted to the environmental mode,
while the remaining half is equally distributed between the two signal output ports,
as illustrated in Fig.~\ref{fig:concept}~(b).
This can be confirmed by substituting $|C|=1$ and $\delta\u{c}=0$ in Eq.~(\ref{eq:S}). 
In contrast, for the symmetric input $\vect{u}\u{in}=\alpha\u{s,in} \vect{u}_{+}$,
the output from the cavity completely vanishes.
This is the phenomenon known as CPA, as illustrated in Fig.~\ref{fig:concept}~(c).
Thus, the CPA process can be switched on and off by the pump light, enabling fast all-optical switching. 
For comparison, the antisymmetric input $\vect{u}\u{in}=\alpha\u{s,in} \vect{u}_{-}$ 
completely suppresses absorption as indicated by $\lambda_-=-1$ and illustrated in Fig.~\ref{fig:concept}~(d).

\section{Experimental setup}
\begin{figure}
 \begin{center}
      \scalebox{1}{\includegraphics{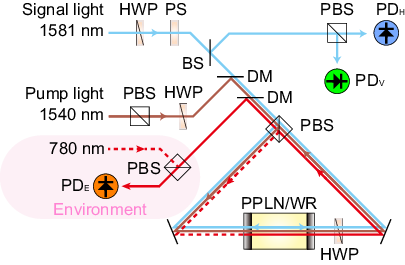}}
    \caption{
      Experimental setup.
      The cyan and red solid lines indicate the signal light at \SI{1581}{nm}
      and frequency-converted light at \SI{780}{nm}, respectively.
      The red dashed line indicates the externally injected environmental light at \SI{780}{nm}.
      Light beams at different wavelengths are combined and separated using dichroic mirrors~(DMs).
      Details are provided in the main text.
    }
 \label{fig:setup}
 \end{center}
\end{figure}
The experimental setup for FC-CPA is shown in Fig.~\ref{fig:setup}. 
We inject a continuous-wave signal light at \SI{1581}{nm} from a tunable laser source~(TSL-570, Santec),
corresponding to $\omega$, into the resonator,
and a continuous-wave pump light at \SI{1540}{nm} generated by a narrow-linewidth laser source~(PLANEX, RIO Lasers),
corresponding to $\omega\u{p}$, used for frequency conversion.
The pump light is amplified by an erbium-doped fiber amplifier~(AEDFA-PM-33-B, Amonics). 
The light at \SI{780}{nm}, corresponding to $\omega\u{e} = \omega + \omega\u{p}$, is regarded as 
the environmental mode interacting with the signal mode through frequency conversion. 

In the experiment, coherent signal injection from both sides is implemented by placing the PPLN/WR inside an interferometer. 
The PPLN/WR, fabricated by NTT Innovative Devices with end-facet coatings provided by Kogakugiken,
forms an FP cavity only for the signal light around \SI{1581}{nm}. This cavity acts as a lossy BS.
On both end faces of the waveguide, dielectric multilayers are coated.
They are designed to have a high reflectance of $\sim$ \SI{94}{\%} at \SI{1581}{nm}. 
The free spectral range~(FSR) of the resonator is approximately \SI{3.5}{GHz} 
considering the length \SI{20}{mm} of the waveguide~\cite{Ikuta2019,Ikuta2022-2}. 
In contrast, the reflectance for \SI{780}{nm} is a few percent
and thus the coating works as anti-reflection. 
The reflectance for the pump light around \SI{1540}{nm} is about \SI{30}{\%} 
which gives no cavity enhancement effect. 

The signal light at \SI{1581}{nm} is decomposed by a polarizing beamsplitter~(PBS)
into horizontally~(H) and vertically~(V) polarized components, 
which then propagate in the clockwise~(cw) and counterclockwise~(ccw) directions 
of an interferometer including the PPLN/WR, respectively. 
The amplitude imbalance and relative phase between the H and V components
are controlled by a half-wave plate~(HWP) and a liquid-crystal phase shifter~(PS), respectively.
In the cw direction, the H-polarized signal light is rotated to V polarization by an HWP inside the interferometer, 
and then coupled into the PPLN/WR. 
The V-polarized light transmitted through the PPLN/WR is detected by a photodetector~($\U{PD_V}$). 
The V-polarized light reflected by the PPLN/WR is converted back to H by the HWP,
and is then detected by another photodetector~($\U{PD_H}$). 
On the other hand, in the ccw direction, the V-polarized signal light is coupled into the PPLN/WR. 
The light transmitted through the PPLN/WR is rotated to H, and detected by $\U{PD_H}$. 
The V-polarized light reflected by the PPLN/WR is detected by $\U{PD_V}$. 

The PPLN/WR satisfies the type-0 quasi-phase-matching condition for frequency conversion,
in which all the relevant fields are V-polarized.
From the phase-matching condition, the \SI{780}{nm} light generated via frequency conversion must propagate
in the same direction as the pump light at \SI{1540}{nm}.
In contrast, the signal light at \SI{1581}{nm} confined in the resonator can contribute to the conversion
regardless of its propagation direction.
For this reason, it is sufficient to use only the pump light in the ccw direction for frequency conversion of the signal light.

\section{Experimental results}
\begin{figure}[t]
 \begin{center}
   \scalebox{0.7}{\includegraphics{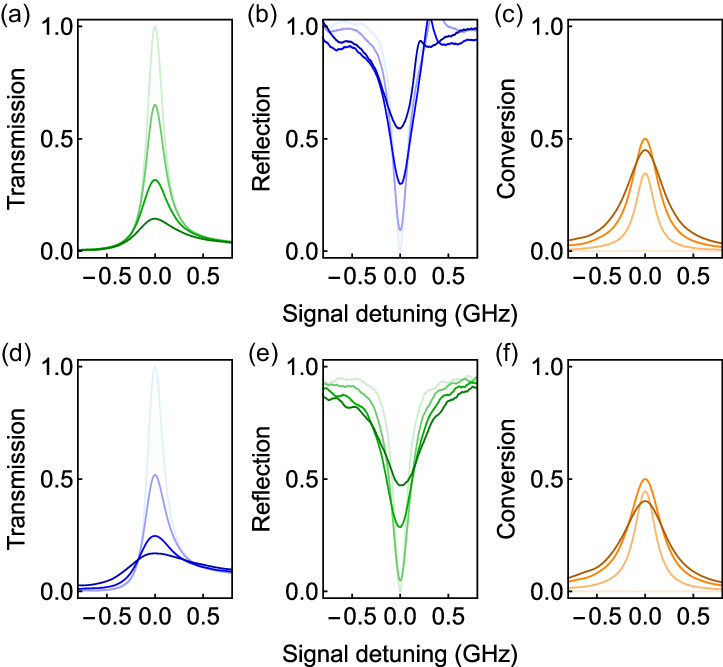}}
   \caption{
     Normalized observed responses from the FC-CPA for input from either cw or ccw direction. 
     The blue, green, and orange colors denote the detector channels $\U{PD}\u{H}$, $\U{PD}\u{V}$, and $\U{PD}\u{E}$,
     respectively, corresponding to the colors used in Fig.~\ref{fig:setup}. 
     Darker colors correspond to higher pump powers. 
     Spectra of (a) transmission, (b) reflection, and (c) conversion are shown for the signal input in the cw direction 
     measured at pump powers of \SI{0}{mW}, \SI{21}{mW}, \SI{72}{mW}, and \SI{145}{mW}.
     (d), (e), and (f) show the corresponding spectra 
     for the ccw direction measured at the pump powers of \SI{0}{mW}, \SI{39}{mW}, \SI{110}{mW}, and \SI{180}{mW}.
   }
 \label{fig:spectra}
 \end{center}
\end{figure}
We first evaluate the output spectra of the FC-CPA while the signal input is injected only from the cw direction. 
The experimental results of the transmission, reflection, and converted-light spectra
measured at $\U{PD}\u{V}$, $\U{PD}\u{H}$, and $\U{PD}\u{E}$ 
for various values of the pump power are shown in Figs.~\ref{fig:spectra}~(a)--(c).
We see that, as the pump power increases, the peak of the transmission spectrum 
and depth of the reflection spectrum decrease monotonically. 
The peak and depth values of the transmission and reflection spectra 
measured without the pump light are taken as the respective maximum values and normalized to 1. 
On the other hand, as expected from Eq.~(\ref{eq:T}),
the peak of the converted-light spectrum first increases with pump power,
reaches a maximum normalized value of 0.5 at \SI{72}{mW}, and then decreases as the pump power is further increased. 
We note that the linewidth of the spectrum increases as the pump power increases. 
This is because the loss from the resonant mode increases, as can be understood from Eq.~(\ref{eq:ac}). 

Similarly, output spectra are observed for signal input only from the ccw direction. 
Figs.~\ref{fig:spectra}~(d)--(f) show the transmission, reflection, and converted-light spectra measured
at $\U{PD}\u{H}$, $\U{PD}\u{V}$, and $\U{PD}\u{E}$, respectively. 
The maximum peak of the converted-light spectrum is obtained at a pump power of \SI{110}{mW},
which is slightly higher than that for signal input from the cw direction. 
Nevertheless, the overall tendency is similar to that observed for the cw direction, 
as is the case for the other spectra.

\begin{figure}[t]
 \begin{center}
   \scalebox{0.7}{\includegraphics{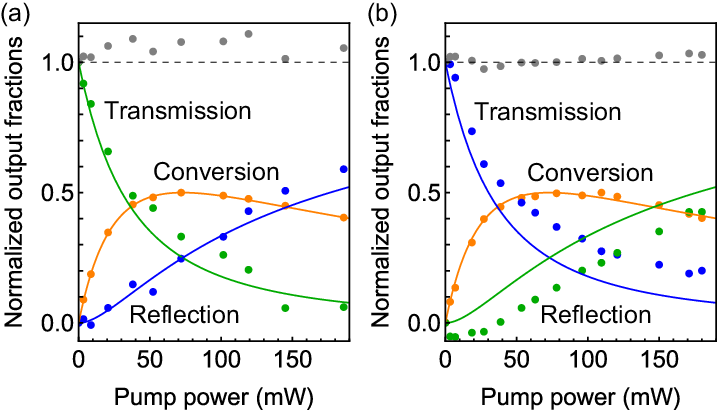}}
   \caption{
     Pump-power dependence of the transmission, reflection, and conversion fractions.
     Blue, green, and orange data were measured at $\U{PD}\u{H}$, $\U{PD}\u{V}$, and $\U{PD}\u{E}$, respectively. 
     The gray points show the sum of the transmission, reflection, and conversion fractions.
     Signal light is input from (a) cw direction, and (b) ccw direction.
   }
 \label{fig:fractions}
 \end{center}
\end{figure}
From the peaks of the transmission and converted-light spectra and the dips of the reflection spectra,
as exemplified in Fig.~\ref{fig:spectra},
we obtain the pump-power dependence of the output fractions from the FC-CPA, as shown in Fig.~\ref{fig:fractions}, 
where the transmission, reflection, and converted-light fractions as well as their sum are plotted. 
We first estimate a parameter $\beta=|C|^2/P$, where $P$ is the pump power. 
To estimate $\beta$, we use the converted-light fraction $2\beta P/(1+\beta P)^2$ derived from Eq.~(\ref{eq:T}), 
because the converted-light channel provides an almost background-free estimate of $\beta$, 
with no pre-existing \SI{780}{nm} signal reaching the detector in the absence of conversion.
This gives nearly identical values of $\beta = \SI{0.014}{mW^{-1}}$ and $\SI{0.013}{mW^{-1}}$ for the cw and ccw directions, respectively, 
indicating comparable frequency-conversion-induced loss in the two directions. 

As shown in Fig.~\ref{fig:fractions},
the theoretical curves calculated using the values of $\beta$ estimated from the converted-light fraction 
capture the overall trends of the transmission, reflection, and converted-light fractions,
although the transmission and reflection data show larger fluctuations. 
The sum of the transmission, reflection, and converted-light fractions remains approximately unity at each pump power,
confirming that the frequency conversion process is well described by Eq.~(\ref{eq:T}). 
Another important observation is that, for single-sided input, the sum of the transmitted and reflected fractions remains
at least 0.5 and never drops to zero. 
This behavior contrasts with the two-sided coherent input case discussed below,
where both output channels of the main system can be simultaneously suppressed. 

\begin{figure}[t]
 \begin{center}
   \scalebox{0.7}{\includegraphics{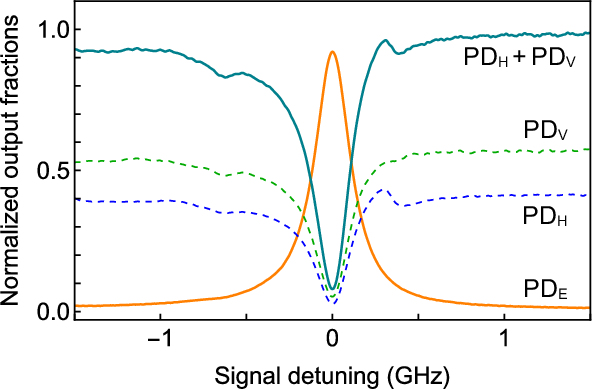}}
   \caption{
     FC-CPA at the pump power of \SI{45}{mW} using signal light incident from both directions. 
     The amount of absorption from the main system~(cyan) into the environment~(orange) reaches approximately \SI{92}{\%},
     which is not achieved when the signal is injected from either the cw or ccw direction alone. 
   }
 \label{fig:cpa}
 \end{center}
\end{figure}
Next, we demonstrate FC-CPA using a superposition of signal light from both directions.
The experiment is performed at a pump power of \SI{45}{mW}.
From Eq.~(\ref{eq:ev}),
the absorption is sensitive to the relative phase between the counterpropagating signal fields. 
Using the PS in Fig.~\ref{fig:setup}, we optimized the relative phase between the H and V components to maximize the absorption.
The experimental result is shown in Fig.~\ref{fig:cpa}. 
The total fractions detected at $\U{PD}\u{H}$ and $\U{PD}\u{V}$ correspond to
the overall energy remaining in the main system at \SI{1581}{nm} after the lossy BS-like transformation in the PPLN/WR. 
Unlike the case of signal injection from only one side,
both output channels detected at $\U{PD}\u{H}$ and $\U{PD}\u{V}$ are simultaneously suppressed at the resonance point
due to energy transfer from the \SI{1581}{nm} main system
to the frequency-converted light at \SI{780}{nm} regarded as the environment.
When the total fraction of the main system in the absence of dissipation under the off-resonant condition is normalized to unity, 
the total dissipation at the resonance point is estimated to be \SI{92}{\%}. 
This value clearly exceeds the maximum dissipation value of 0.5 obtained for signal injection from a single side, 
indicating the successful observation of FC-CPA.

\begin{figure}[t]
 \begin{center}
   \scalebox{0.7}{\includegraphics{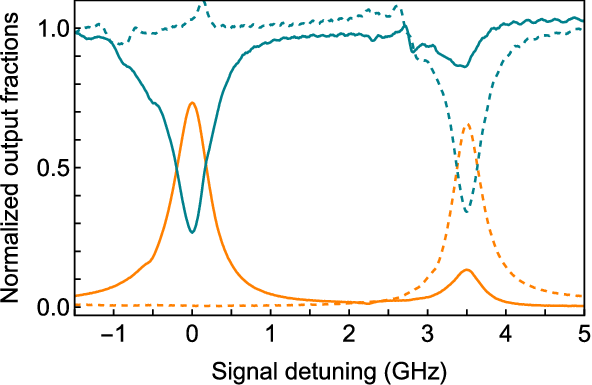}}
   \caption{
     Relative-phase dependence of absorption for inputs from both cw and ccw directions. 
     The cyan and orange curves indicate the amount of energy in the main system and the environment, respectively. 
     The solid and dashed curves are
     for the input light of $\vect{u}\u{in}=\alpha\u{s,in}\vect{u}_{+}$ and $\alpha\u{s,in}\vect{u}_{-}$ at zero detuning. 
     For reference, the absorption behavior at the adjacent resonance point at 3.5-GHz detuning is also shown. 
   } 
 \label{fig:phase}
 \end{center}
\end{figure}
We then investigated the dependence of absorption on the relative phase between the two signal inputs. 
The solid curve in Fig.~\ref{fig:phase} shows the experimental result of FC-CPA at a pump power of \SI{186}{mW}. 
Although this pump power exceeds the estimated critical value for CPA, 
a high absorption of \SI{73}{\%} is still observed at zero detuning from resonance for input light 
with an optimal relative phase of $0$.
On the other hand, when the relative phase of the input light is tuned to $\pi$,
the absorption is completely suppressed at zero detuning, as shown by the dotted curve. 
For comparison, we also show the response at the adjacent resonance point, separated by the FSR of \SI{3.5}{GHz}.
The absorption differs from that at zero detuning 
because the interferometer introduces a frequency-dependent relative phase between the two counterpropagating inputs. 
Thus, the same phase setting does not correspond to the same input superposition at the adjacent resonance.

Finally, we perform an experiment to control absorption in the main system 
via energy injection from the environment. 
In the previous experiments, light was injected only from the main system as in a standard CPA setup. 
However, by simultaneously injecting an environmental field with an appropriate phase and intensity, 
it becomes possible to either enhance or suppress the energy absorption from the main system. 
As described around Eq.~(\ref{eq:ev}), 
the conventional CPA condition without the input of $a\u{e}$ is achieved only for $|C|=1$ and $\delta\u{c} = 0$. 
In contrast, an input to the environmental mode $a\u{e}$ can compensate for deviations from the conventional CPA condition, 
such as $|C| \neq 1$ and $\delta\u{c} \neq 0$. 
Specifically, we consider an input state as $\vect{v}\u{in} = (\alpha\u{s,in}/\sqrt{2}, \alpha\u{s,in}/\sqrt{2}, \alpha\u{e,in})^{\mathsf T}$. 
In this case, using Eq.~(\ref{eq:T}),
the output state is described by $\vect{v}\u{out}=(\alpha\u{1,out}, \alpha\u{2,out}, \alpha\u{e,out})^{\mathsf T}$, where 
\begin{align}
  \alpha\u{1,out} = \alpha\u{2,out} &= \frac{1}{\sqrt{2}A}((1-A)\alpha\u{s,in} + C^* \alpha\u{e,in}),
              \label{eq:asout}\\
  \alpha\u{e,out} &= \frac{1}{A}(-C\alpha\u{s,in} + (1-A^*)\alpha\u{e,in}).
              \label{eq:aeout}
\end{align}
From the equation, by tuning the complex amplitude of $\alpha\u{e,in}$, the energy remaining in the main system 
can be decreased or increased compared with the case $\alpha\u{e,in}=0$, 
due to interference between the signal and environmental light. 
In particular,
$\alpha\u{e,in}/\alpha\u{s,in}=(A-1)/C^*$ and $\alpha\u{e,in}/\alpha\u{s,in}=(2A-1)/C^*$ yield 
$\alpha\u{1,out}=\alpha\u{2,out}=0$ and 
$\alpha\u{1,out}=\alpha\u{2,out}=\alpha\u{s,in}/\sqrt{2}$, respectively. 
This shows that the absorption in the main system can be enhanced or suppressed
by the environmental input over a broad range of $C$ and $\delta\u{c}$,
although larger environmental input is required away from the resonance and critical coupling. 

\begin{figure}[t]
 \begin{center}
   \scalebox{0.70}{\includegraphics{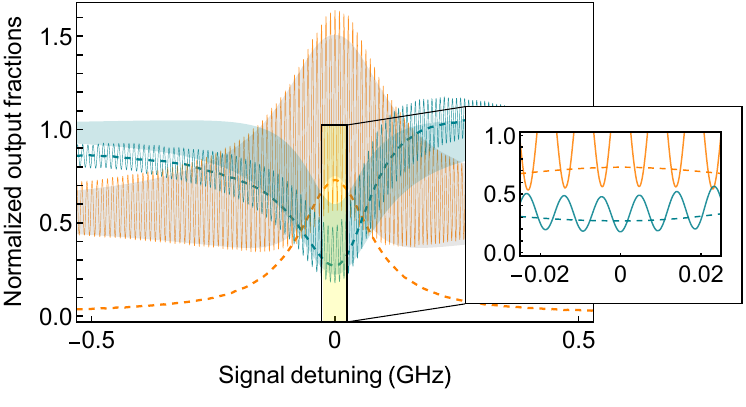}}
   \caption{
     Enhancement and suppression of absorption via energy injection from the environment. 
     The cyan and orange curves indicate the amount of energy in the main system and the environment, respectively.
     The inset magnifies the oscillatory structure around zero detuning,
     which effectively corresponds to scanning the relative phase of the injected environment field. 
     The dashed curves without oscillations show the conventional absorption behavior in the absence of the \SI{780}{nm} input.
     The theoretical predictions for the envelopes of the \SI{1581}{nm} and \SI{780}{nm} output fields
       are depicted by the filled regions with cyan and orange, respectively. 
   } 
   \label{fig:gain}
 \end{center}
\end{figure}
For the proof-of-concept experiment, we prepared \SI{780}{nm} light 
using another PPLN waveguide~(not shown in Fig.~\ref{fig:setup})
by sum frequency generation of the \SI{1540}{nm} pump and \SI{1581}{nm} signal light. 
Starting from the FC-CPA configuration without environmental input at a pump power of \SI{12}{mW}, 
we additionally inject \SI{780}{nm} light from the ccw direction.
The experimental result is shown in Fig.~\ref{fig:gain}.
The cyan and orange solid lines are the normalized outputs of the main system and environment, respectively,
in the presence of the environmental input whereas the dashed lines represent outputs without the environmental input.
The oscillation seen on the solid lines is specific to the present experimental implementation 
and does not reflect an intrinsic feature of the underlying mechanism.
As mentioned above, the \SI{780}{nm} light is generated using the \SI{1581}{nm} signal light to ensure phase coherence.
This configuration, however, couples the phase of the \SI{780}{nm} light to the detuning of the \SI{1581}{nm} signal.
Owing to the long optical path length of the \SI{780}{nm} light in the present setup,
this effect is amplified, resulting in the observed oscillatory behavior.
Therefore, the upper and lower envelopes of the main system output~(solid cyan)
in comparison with the output without the environmental input~(dashed cyan) should be interpreted
as the indicator for the magnitude of suppression and enhancement assisted by the environmental input.
Then as seen in the inset of Fig.~\ref{fig:gain},
the absorption is clearly either enhanced or suppressed depending on the detuning. 

The behavior of the envelopes is qualitatively reproduced by extending the input-output model
described by Eqs.~(\ref{eq:T}), (\ref{eq:asout}), and (\ref{eq:aeout}) 
to include the finite mode-matching efficiency of the injected environmental field 
and an effective interference contrast in the \SI{1581}{nm} output~(see Appendix).
In this model,
the mode-matched component of the \SI{780}{nm} input coherently participates in the frequency-conversion process,
whereas the unmatched component is added as an incoherent contribution to the measured environment intensity. 
The model explains the behavior of the environmental output well, 
including the vertical offset as seen in Fig.~\ref{fig:gain}. 

On the other hand, the experimentally observed contrast of the envelopes in the \SI{1581}{nm} output is 
smaller than expected from the finite mode-matching alone. 
This reduced contrast is phenomenologically taken into account in the model
through an effective interference contrast $V$ in Eq.~(\ref{eq:Isout}) in Appendix.
We speculate that the reduced contrast is caused by high sensitivity of the \SI{1581}{nm} output field to the relative phases
between the transmission and reflection of the cavity in the interferometer~(note that
by contrast the \SI{780}{nm} field only passes through the cavity). 
Achieving more complete control of absorption would require further optimization of the complex amplitude of the environmental input,
together with improved mode-matching of the injected environment field and interference contrast in the \SI{1581}{nm} output.
Nevertheless, the result demonstrates that optical injection from the environmental system
can contribute to both enhancement and suppression of energy absorption in the main system.

The observed environment-assisted control of absorption can be interpreted
as an interference mechanism similar to electromagnetically induced transparency~(EIT)~\cite{Fleischhauer2005}. 
In EIT, dark and bright states arise from superpositions of two ground states 
coupled to a common excited state by the probe and control fields. 
By contrast, in FC-CPA, the dark and bright modes are formed from the two input channels of the main system.
This structure introduces a control degree of freedom distinct from the conventional EIT, 
allowing the dissipation of the symmetric supermode to be coherently controlled by the environmental field,
whereas the antisymmetric supermode remains decoupled from the loss channel.

The environment-assisted framework also provides an alternative perspective on CPA. 
While conventional CPA is often described in terms of a non-unitary scattering matrix 
with a zero eigenvalue after eliminating inaccessible loss channels~\cite{Chong2010,Baranov2017}, 
our system realizes the effective loss through coherent coupling to an accessible frequency-converted mode.
Thus, the non-unitary CPA dynamics of the main system can be viewed as the projection of a higher-dimensional coherent mode transformation onto the main-system subspace.  
The environmental mode is not merely a loss reservoir, but a controllable optical port. 

 \section{Conclusion}
In conclusion, we have proposed and experimentally demonstrated FC-CPA in a PPLN/WR,
achieving up to \SI{92}{\%} absorption under coherent two-sided excitation. 
Our results show that the nonlinear cavity acts as a coherently tunable dissipative element,
with the effective loss all-optically controlled via the pump field rather than intrinsic material absorption. 
Because the frequency-conversion process realizes a linear optical interaction between the system and environmental modes, FC-CPA can, in principle, be extended directly to single-photon inputs with the same coherent absorption mechanism as for classical light. 
Moreover, by injecting an optical field into the accessible environmental mode,
we demonstrate that dissipation in FC-CPA is not a fixed property of the system but a coherently controllable degree of freedom.
This environment-assisted control provides a new route to coherent manipulation of dissipation
and opens pathways toward reconfigurable photonic interfaces, 
including phase-dependent optical routing and switching for classical and quantum photonic networks. 

\section*{Acknowledgements}
The authors acknowledge discussions at the 6th and 11th QUATUO meetings 
(QUAntum Theory and Technology Unofficial meetings), held in Japan in 2017 and 2025.
R.I. is grateful to Kurama Hirano for assistance with the experiment. 
This work was supported by JST FOREST Program JPMJFR222V,
Asahi Glass Foundation,
R \& D of ICT Priority Technology Project JPMI00316,
and MEXT/JSPS KAKENHI JP25K01263. 

\appendix

\section{Theoretical model for Fig.~\ref{fig:gain}}

To explain the results shown in Fig.~\ref{fig:gain}, we extend the input-output model 
to include finite mode matching of the injected environment field.
For simplicity, we take $C$ to be real in the fitting model,
since the phase of $C$ can be included in the definition of the complex amplitude of the injected environment field $\alpha\u{e,in}$.
We write the mode-matched component of the environment input as 
$\sqrt{\eta}\alpha\u{e,in}$, where $0 \leq \eta \leq 1$.
The remaining component is assumed not to participate in the coherent frequency-conversion process.

From Eq.~(\ref{eq:aeout}),
the coherent component of the environment output is described by
\begin{align}
  \alpha\u{e,out}^{\U{(coh)}}
  =
  \frac{1}{A}
  \left(
  -C\alpha\u{s,in}+(1-A^*)\sqrt{\eta}\alpha\u{e,in}
  \right).
\end{align}
The intensity in the environment output is then modeled as
\begin{align}
 I\u{e}
 &=
 \frac{1}{|A|^2}
 \left|
 -C\alpha\u{s,in}+(1-A^*)\sqrt{\eta}\alpha\u{e,in}
 \right|^2
   +(1-\eta)|\alpha\u{e,in}|^2 .
   \label{eq:Ieout}
\end{align}
Here, the last term represents the unmatched component of the injected \SI{780}{nm} field,
which contributes as an incoherent background in the environment output.
The upper and lower boundaries of the theoretical band are obtained by choosing the phase of $\alpha\u{e,in}$
such that the interference term is maximally constructive or destructive. 

On the other hand, from Eq.~(\ref{eq:asout}),
each output channel of the main system includes the mode-matched component of the environmental mode as 
\begin{align}
  \alpha\u{1,out}^{\U{(coh)}} =
  \alpha\u{2,out}^{\U{(coh)}} =
  \frac{1}{\sqrt{2}A}
  \left(
  (1-A)\alpha\u{s,in} + C\sqrt{\eta}\alpha\u{e,in}
  \right).
\end{align}
Because the \SI{1581}{nm} output involves interference between the original signal field and the environment-induced component,
it is more sensitive to residual interferometric phase, polarization, and normalization imperfections than the \SI{780}{nm} field. 
We therefore introduce an effective interference contrast $V$ in the interference term. 
The corresponding intensity is written as
\begin{align}
  I\u{s} &= |\alpha\u{1,out}^{\U{(coh)}}|^2 + |\alpha\u{2,out}^{\U{(coh)}}|^2\\
  &= 
 \frac{1}{|A|^2}
 \left(
   |1-A|^2|\alpha\u{s,in}|^2 + C^2\eta |\alpha\u{e,in}|^2
   \right.
 \nonumber\\
 &\hspace{2.5em}
\left. + 2V\,\U{Re}
 \left\{
 (1-A^*) C \sqrt{\eta}\alpha\u{s,in}^*\alpha\u{e,in}
 \right\}
   \right)
   \label{eq:Isout}
\end{align}

As shown in Fig.~\ref{fig:gain}, the signals detected in the main system are distorted by the interferometric response of the transmitted and reflected fields from the cavity.
This effect makes a direct quantitative fit of the \SI{1581}{nm} output unreliable. 
In contrast, the converted-light signal at \SI{780}{nm} is less affected by this interferometric distortion and provides a cleaner measure of the frequency-conversion process.
We therefore first estimate the parameters $C$ and $\gamma$ 
from the data obtained without the injected \SI{780}{nm} environment field. 
The estimated values are $C^2=0.30$ and $\gamma=\SI{138}{MHz}$. 

We then fit the \SI{780}{nm} output measured with the environment input using Eq.~(\ref{eq:Ieout}),
including the finite mode-matching efficiency $\eta$ of the injected environment field. 
To avoid the rapid oscillatory component caused by the experimental phase configuration,
the fit is performed to the envelope of the observed \SI{780}{nm} signal. 
This procedure gives $\eta=0.32$ and $|\alpha\u{e,in}/\alpha\u{s,in}|^2=1.0$. 
As shown in Fig.~\ref{fig:gain}, 
the resulting theoretical band captures the overall envelope of the experimental \SI{780}{nm} output. 

Using the same values of $C$, $\gamma$, and $\eta$,
we then compare the model with the \SI{1581}{nm} output in the main system.
We use a representative value of $V=0.50$ to reproduce the observed contrast between the upper and lower envelopes. 
From Fig.~\ref{fig:gain}, with this value, the model captures the main features of the \SI{1581}{nm} data,
including both enhancement and suppression of absorption.

%%%%%%%%%% If preparing manually:

\end{document}